\documentclass{ws-procs975x65}

\begin{document}

\title{RELATIVISTIC FIGURES OF EQUILIBRIUM:\\ 
FROM MACLAURIN SPHEROIDS TO KERR BLACK HOLES}

\author{REINHARD MEINEL}

\address{University of Jena, Theoretisch-Physikalisches Institut,\\
Max-Wien-Platz 1, 07743 Jena, Germany\\
\email{meinel@tpi.uni-jena.de}}

\begin{abstract}
Analytical and numerical results on equilibrium configurations of rotating fluid bodies within Einstein's theory of gravitation are reviewed. Particular emphasis is placed on continuous parametric transitions to black holes. In this connection the uniqueness of extremal Kerr black holes is discussed. 

\end{abstract}

\bodymatter

\section{Introduction}
The theory of figures of equilibrium of rotating fluid bodies under the influence of their own gravitational field is relevant for understanding the shape and the  structure of planets and stars. The problems to be solved are rather involved even within Newton's theory of gravitation, and analytic solutions are scarce. Basic references are the books by Lichtenstein\cite{li} and Chandrasekhar\cite{ch}. Within Einstein's theory of gravitation, which is relevant for describing compact objects like neutron stars, analytic solutions are only available in certain limiting cases. In general, a numerical treatment is necessary. For a detailed account of relativistic figures of equilibrium see Ref.~\refcite{rfe} and the book by Friedman and Stergioulas\cite{fs}. In the following, a brief review of some analytical and numerical results will be given.
 
\section{Analytically tractable limiting cases}
Analytic solutions are known in the Newtonian limit (e.g.~the Maclaurin spheroids) and in the non-rotating limit (e.g.~the Schwarzschild spheres of constant mass-density). In addition, the problem of a uniformly rotating, infinitesimally thin disc of dust was analytically solved within Einstein's theory of gravitation\cite{nm1,nm2,nm3}. This was possible by applying the ``inverse scattering method'' to the corresponding boundary value problem of the vacuum Einstein equations\cite{nm4}. A very accurate approximate solution to this problem had already been presented by Bardeen and Wagoner\cite{bw}. Note that the Newtonian limit of the relativistic disc solution coincides with the disc limit of the Maclaurin spheroids. The extreme relativistic limit leads to a maximally rotating Kerr black hole, see Sec.~\ref{bh}.

\section{Numerical treatment of the general case}
Uniformly rotating relativistic figures of equilibrium are necessarily symmetric about the axis of rotation as deviations from axisymmetry would lead to gravitational radiation. Applying the model of a (cold) perfect fluid, one has to specify an equation of state $\mu=\mu(p)$ relating mass(-energy)-density $\mu$ and pressure $p$. For a given equation of state the parameter space of the equilibrium solutions is two-dimensional. For example, one can choose the total (gravitational) mass $M$ and the angular momentum $J$ as the two parameters. A complete overview of the relativistic solution space for the equation of state $\mu={\rm constant}$ was given in Ref.~\refcite{afkmps}. The results were obtained by a multi-domain (pseudo-)spectral method\cite{akm1} that is applicable to arbitrary equations of state. Note that the solutions include the ``relativistic Dyson rings''\cite{akm2}. An interesting general feature of the exterior field of uniformly rotating perfect fluid bodies seems to be that all gravitational multipole moments beyond $M$ and $J$, in particular the quadrupole moment, are greater than or equal to the corresponding moments of the Kerr metric with the same mass and angular momentum\cite{fk}. The equality holds (presumably only) in the case of a ``black hole limit'' where all multipole moments become precisely those of the extreme Kerr metric\cite{m1}. This will be discussed in more detail in the next section.

\section{Transition to black holes}\label{bh}
It was shown in Ref.~\refcite{m2} that a parametric transition of a uniformly rotating perfect fluid configuration to a black hole occurs if and only if the relation 
\begin{equation}
Mc^2-2\Omega J \to 0
\label{bhl}
\end{equation}
is satisfied when approaching the limit, where $\Omega$ is the angular velocity of the fluid body and $c$ denotes the speed of light. In the limit, $\Omega$ becomes identical with the ``angular velocity of the horizon''. Together with the uniqueness of the Kerr black holes, Eq.~(\ref{bhl}) then implies that it is always an extremal Kerr black hole that results, i.e.~the one with the maximal angular momentum $J=GM^2/c$ for a given mass ($G$ denotes Newton's gravitational constant). This conclusion, however, needs an extension of the Kerr uniqueness results\cite{car1,car2,rob,cc} to the case with a degenerate horizon. As already mentioned in Ref.~\citen{m2} it can indeed be shown that the only asymptotically flat, stationary and axisymmetric black hole solutions with a single degenerate (Killing) horizon surrounded by a vacuum, are given by the extremal Kerr black holes. A corresponding proof was published in Ref.~\citen{rfe} (end of section 2.4). Recently, two independent proofs appeared\cite{ahmr,fl}.

The relativistic disc solution\cite{nm3} exists for $0<GM^2/cJ\le 1$. For $GM^2/cJ \to 1$ a black hole limit is reached. Details can be found in Ref.~\refcite{m3}. Convincing numerical evidence for parametric transitions to black holes was also presented for the relativistic Dyson rings\cite{akm2} and ring solutions with other equations of state\cite{fha,lpa}. It should be noted that a characteristic ``separation of spacetimes'' occurs in the parameter limit: From the ``exterior point of view'', the extreme Kerr metric outside the horizon results. From the ``interior point of view'', a regular, non-asymptotically flat spacetime with the extreme Kerr ``near-horizon geometry'' at spatial infinity results. Similar phenomena were observed for limiting solutions to the static, spherically symmetric Einstein--Yang--Mills--Higgs and Einstein--Maxwell equations\cite{bfm,lw,lz}.

\section*{Acknowledgments}
This research was supported by the Deutsche Forschungsgemeinschaft (DFG) through the SFB/TR7 ``Gravitations\-wellenastronomie''.

\end{document}